# Ternary Policy Iteration Algorithm for Nonlinear Robust Control

Jie Li, Shengbo Eben Li*, Yang Guan, Jingliang Duan, Wenyu Li, Yuming Yin

*Abstract*—The uncertainties in plant dynamics remain a challenge for nonlinear control problems. This paper develops a ternary policy iteration (TPI) algorithm for solving nonlinear robust control problems with bounded uncertainties. The controller and uncertainty of the system are considered as game players, and the robust control problem is formulated as a two-player zero-sum differential game. In order to solve the differential game, the corresponding Hamilton-Jacobi-Isaacs (HJI) equation is then derived. Three loss functions and three update phases are designed to match the identity equation, minimization and maximization of the HJI equation, respectively. These loss functions are defined by the expectation of the approximate Hamiltonian in a generated state set to prevent operating all the states in the entire state set concurrently. The parameters of value function and policies are directly updated by diminishing the designed loss functions using the gradient descent method. Moreover, zero-initialization can be applied to the parameters of the control policy. The effectiveness of the proposed TPI algorithm is demonstrated through two simulation studies. The simulation results show that the TPI algorithm can converge to the optimal solution for the linear plant, and has high resistance to disturbances for the nonlinear plant.

## I. INTRODUCTION

The robust control theory has been continuously investigated since it was originally formulated by Zames [1] in 1981, which can handle model parameter perturbations or external disturbances in linear control problems [2]-[4]. Recent years have witnessed enormous applications of robust control in the industrial field, such as aircraft control and vehicle platoon control with uncertain dynamics or communication delay [5]-[7]. The essential point of dealing with nonlinear uncertain dynamics is to solve the Hamilton-Jacobi-Isaacs (HJI) equation [8], [9], which is a nonlinear partial differential equation. Compared with the well-known Hamilton-Jacobi-Bellman (HJB) equation in optimal control problem, the utility function of the HJI equation is non-positive definite due to the existence of perturbation or disturbance, so it tends to be more challenging to solve analytically.

Reinforcement learning (RL) is an evolving computational method to iteratively learn a control policy via interacting with an environment [14]-[16], which can solve the HJI equation in an iterative manner. Two commonly used RL methods are value iteration and policy iteration, and value iteration can be regarded as a special case of policy iteration [17]. The policy iteration algorithm consists of two alternating steps, namely, policy evaluation and policy improvement. The former step updates the value function of the given control policy, and the latter step generates a greedy policy using the updated value function [16].

The concept of policy iteration has been applied to solve nonlinear $H_\infty$ control problems in the last few years [18]. Abu-Khalaf *et al.* [19] extended the policy iteration techniques to find the optimal control policies for the $H_\infty$ suboptimal nonlinear control problems with affine and constrained control inputs. In order to encode the constrained inputs, a special quasi-norm was used to formulate a non-quadratic zero-sum game, and then the HJI equation was decomposed into a sequence of differential equations and solved by policy iterations under the control and disturbance inputs. Abu-Khalaf *et al.* [20] further employed neural network (NN) approximate functions as well as inner- and outer-loop policy iterations to solve the zero-sum game of affine systems with control saturation. Specifically, the disturbance policy and matching value function were updated in the inner loop, and the control policy was updated in the outer loop. However, a stable initial control policy was required for uniform convergence. Al-Tamimi *et al.* [21] proposed an online model-free iterative algorithm relying on Q-learning for linear discrete-time zero-sum game. The action dependent value function was solved via satisfying the corresponding game algebraic Riccati equation (GARE). Two action networks and one critic network were obtained using adaptive critic methods.

Methods mentioned above were established on the assumptions that the saddle points exist, the dynamic system is fully known and the initial control policy is stabilizing. In order to overcome these limitations, Zhang *et al.* [22] proposed a policy iteration method which was stable and convergent in the presence or absence of saddle points for zero-sum differential games. Furthermore, an NN-based online simultaneous policy update algorithm was developed in [23] to solve the HJI equation for the affine nonlinear system without knowledge of internal system dynamics, which was embedded in the online measurement of the states and evaluation of the loss function. The convergence property was established by proving that the algorithm was mathematically equivalent to finding a fixed point in Newton's method. For entirely unknown dynamic systems, an online algorithm building on an off-policy integral version of the Bellman equation was proposed to design an $H_\infty$ tracking controller in [24]. Besides, Liu *et al.* [25] designed a robust adaptive control algorithm for nonlinear systems with uncertainties, which relaxed the requirement of initial stabilizing control policy using only one critic NN. However, all states in the entire state set needed to be handled separately and only polynomial bases

This work is supported by International Science & Technology Cooperation Program of China under 2019YFE0100200, and Beijing Natural Science Foundation with JQ18010. All correspondence should be sent to S. Li with email: lisb04@gmail.com.

J. Li, S. Li, Y. Guan, J. Duan, W. Li, Y. Yin are with School of Vehicle and Mobility, Tsinghua University, Beijing, 100084, China. E-mail: ({jie-li18, guany17, djl15}@mails.tsinghua.edu.cn, {lisb04, yinyuming89}@gmail.com, liwenyu@mail.tsinghua.edu.cn)

could be selected as neurons in those offline algorithms, which made them less practical.

In order to overcome the above limitations and improve the practicability and versatility of the algorithm, this paper proposes a single-loop reinforcement learning algorithm, called ternary policy iteration (TPI), which has potential to handle non-affine nonlinear dynamics for robust optimal control problem with zero-initialized control policy. Our main contributions include: (1) the expectation values in the generated state set rather than all states in the entire state set need to be operated during the learning process to improve the practicability; (2) both polynomials and fully connected networks can be employed to approximate value function and policies, which improves the generality of the algorithm.

The rest of paper is organized as follows. In Section II, descriptions of the considered nonlinear robust control problem and differential game are given, and the corresponding HJI equation is derived. Section III introduces the proposed TPI algorithm. Two simulation scenarios are presented in section IV to demonstrate the effectiveness and robustness of the proposed algorithm. A brief conclusion is given in section V.

## II. PROBLEM DESCRIPTION

Consider the continuous-time affine nonlinear plant

$$\dot{x} = f(x) + g(x)u + k(x)w \quad (1)$$

with the state $x \in \mathbb{R}^n$, the control input $u \in \mathbb{R}^m$, and the uncertainty $w \in \mathbb{R}^q$ caused by modelling error or external disturbance. $f(x) \in \mathbb{R}^n$, $g(x) \in \mathbb{R}^{n \times m}$ and $k(x) \in \mathbb{R}^{n \times q}$ are nonlinear differentiable on a state set $\Omega$ containing the origin.

The $H_\infty$ norm of the closed-loop transfer function from the disturbance input $w$ to the objective output $z$ is defined as

$$\|T_{zw}\|_\infty^2 = \sup_w \frac{\|z\|_2^2}{\|w\|_2^2} \quad (2)$$

where $z = \begin{bmatrix} \sqrt{Q}x \\ \sqrt{R}u \end{bmatrix} \in \mathbb{R}^{n+m}$, $Q \geq 0 \in \mathbb{R}^{n \times n}$, $R > 0 \in \mathbb{R}^{m \times m}$.

The uncertainty $w$ is assumed to be bounded by objective output, i.e. $w = \Delta \cdot z$, and $\|\Delta\|_\infty \leq 1/\gamma$. It can be known from the small gain theorem [26] that the closed-loop system is stable if $\|T_{zw}\|_\infty \|\Delta\|_\infty < 1$, i.e.

$$\sup_w \frac{\|z\|_2^2}{\|w\|_2^2} < \gamma^2 \quad (3)$$

which is known as the $H_\infty$ suboptimal control problem.

In order to reduce the closed-loop gain and increase stability margin of the plant, the $H_\infty$ optimal control problem aims to minimize the $H_\infty$ norm of the closed-loop transfer function by changing the control input $u$, i.e.

$$\min_u \|T_{zw}\|_\infty^2 \quad (4)$$

### A. Transform Robust Control to Zero-Sum Game

With the definition of $H_2$ norm, $H_\infty$ suboptimal control problem (3) can be rewritten as

$$\int_t^\infty (z^T z - \gamma^2 w^T w) \, d\tau < 0, \forall w \in L_2[t, \infty) \quad (5)$$

where $t$ is the initial time of the control problem.

Note that $z^T z = x^T Q x + u^T R u$, the value function $V(x)$ or the performance function $J(u, w)$ can be defined as

$$V(x) = J(u, w) = \int_t^\infty l(x, u, w) \, d\tau \quad (6)$$

where $l(x, u, w) = x^T Q x + u^T R u - \gamma^2 w^T w$ is the utility function.

Take partial derivative of $t$ for (6), and an equation of the Hamiltonian is formed as

$$H\left(x, u, w, \frac{\partial V(x)}{\partial x}\right) = x^T Q x + u^T R u - \gamma^2 w^T w + \frac{\partial V(x)}{\partial x^T}(f(x) + g(x)u + k(x)w) = 0 \quad (7)$$

which is widely applied in the design of control algorithms.

According to the relationship between $H_\infty$ control and zero-sum game [26], the solvability of the $H_\infty$ optimal control problem (4) is equivalent to that of the two-player zero-sum differential game, which can be expressed as

$$V^*(x) = J(u^*, w^*) = \min_u \max_w J(u, w) \quad (8)$$

where $V^*(x)$ is the optimal value function and $J(u^*, w^*)$ is the optimal performance function. Next, we will show how to solve this zero-sum game.

### B. Hamilton-Jacobi-Isaacs (HJI) Equation

The two-player zero-sum differential game (8) has a unique solution if a saddle point exists [27], i.e.

$$\min_u \max_w J(u, w) = \max_w \min_u J(u, w) \quad (9)$$

or

$$J(u^*, w) \leq J(u^*, w^*) \leq J(u, w^*), \forall u, w \in L_2[t, \infty) \quad (10)$$

which is the well-known Nash condition. The optimal solution $(u^*, w^*)$ is called saddle point, where $u^*$ and $w^*$ are at equilibrium, neither of which has a motivation to change to make its performance better.

A necessary condition for Nash condition is Isaacs' condition [27], which can be seen as an extension of Pontryagin maximum principle, i.e.

$$\min_u \max_w H\left(x, u, w, \frac{\partial V(x)}{\partial x}\right) = \max_w \min_u H\left(x, u, w, \frac{\partial V(x)}{\partial x}\right) \quad (11)$$

or

$$H\left(x, u^*, w, \frac{\partial V(x)}{\partial x}\right) \leq H\left(x, u^*, w^*, \frac{\partial V(x)}{\partial x}\right) \leq H\left(x, u, w^*, \frac{\partial V(x)}{\partial x}\right), \forall u, w \in L_2[t, \infty) \quad (12)$$

The following subsections will derive the HJI equation for the nonlinear case and linear case, respectively.

*1) HJI Equation in Nonlinear Case*

For Hamiltonian in Isaacs' condition, applying stationarity conditions $\partial H / \partial u = 0$ and $\partial H / \partial w = 0$ gives

$$u = \arg\min_u H\left(x, u, w, \frac{\partial V(x)}{\partial x}\right) = -\frac{1}{2} R^{-1} g^T(x) \frac{\partial V(x)}{\partial x} \quad (13)$$

$$w = \arg\max_w H\left(x, u, w, \frac{\partial V(x)}{\partial x}\right) = \frac{1}{2\gamma^2} k^T(x) \frac{\partial V(x)}{\partial x} \quad (14)$$

Since $\partial^2 H/\partial u^2 = 2R > 0$ and $\partial^2 H/\partial w^2 = -2\gamma^2 < 0$, (13) and (14) together forms a stationary point, which satisfies Isaacs' condition (12). Substituting them to (7) yields the HJI equation of nonlinear dynamics

$$\begin{aligned} x^T Q x + \frac{\partial V(x)}{\partial x^T} f(x) - \frac{1}{4} \frac{\partial V(x)}{\partial x^T} g(x) R^{-1} g^T(x) \frac{\partial V(x)}{\partial x} \\ + \frac{1}{4\gamma^2} \frac{\partial V(x)}{\partial x^T} k(x) k^T(x) \frac{\partial V(x)}{\partial x} = 0 \end{aligned} \quad (15)$$

The positive semi-definite solution $V^*(x)$ to the HJI equation is the optimal value of the zero-sum game (8). Thus, (15) can be written as

$$\min_u \max_w H\left(x, u, w, \frac{\partial V^*(x)}{\partial x}\right) = 0 \quad (16)$$

The boundary condition of the HJI equation is $V(x_e) = 0$, where $x_e$ is an equilibrium state which is usually a zero vector in a state regulator problem. Moreover, assuming that the value function $V(x)$ is continuously differentiable, (16) can be obtained by applying Bellman's principle of optimality [20], which is similar to the derivation of the HJB equation in optimal control problem.

*2) HJI Equation in Linear Case*

For a linear system $\dot{x} = Ax + Bu + Dw$, suppose the value function is a quadratic form of the state, i.e. $V(x) = x^T P x$, where $P$ is symmetric positive. Applying two stationarity conditions again gives

$$u = \arg\min_u H\left(x, u, w, \frac{\partial V(x)}{\partial x}\right) = -R^{-1} B^T P x \quad (17)$$

$$w = \arg\max_w H\left(x, u, w, \frac{\partial V(x)}{\partial x}\right) = \frac{1}{\gamma^2} D^T P x \quad (18)$$

Then, the HJI equation reduces to the following game algebraic Riccati equation (GARE) for the linear system [27]

$$A^T P + PA + Q - P(BR^{-1}B^T - \gamma^{-2}DD^T)P = 0 \quad (19)$$

whose solution can be used to verify the correctness of the algorithm in this study.

In conclusion, the HJI equation (15) of the nonlinear system is a nonlinear partial differential equation, whose analytical solution is almost impossible to seek, while the GARE of the linear system is relatively easy to solve. In the next section, a policy iteration algorithm will be proposed to find a numerical solution to the HJI equation.

III. TERNARY POLICY ITERATION ALGORITHM

In this section, we will introduce our ternary policy iteration (TPI) algorithm, which contains three designed loss functions and three update phases. We will first discuss the deficiencies of existing methods, e.g., updating value function and policies in the entire state set is impractical, or algorithms are not applicable to non-affine systems [19], [23]. Then, the details of our algorithm will be presented.

*A. Existing Robust Policy Iteration Algorithms*

Policy iteration is an iterative method widely used in RL, which involves alternating circulations between policy evaluation and policy improvement, so as the existing robust policy iteration algorithms.

In the policy evaluation step, the control policy $u^k$ and the disturbance policy $w^k$ are fixed, and the value function $V^k(x)$ is updated by solving the following differential equation for $\forall x \in \Omega$

$$\begin{aligned} H\left(x, u^k, w^k, \frac{\partial V^k(x)}{\partial x}\right) = x^T Q x + u^{k^T} R u^k \\ -\gamma^2 w^{k^T} w^k + \frac{\partial V^k(x)}{\partial x^T} f(x, u^k, w^k) = 0 \end{aligned} \quad (20)$$

In policy improvement steps, the value function $V^k(x)$ is given to improve policies $u^{k+1}$ and $w^{k+1}$ by minimizing or maximizing corresponding Hamiltonian $H\left(x, u^k, w^k, \frac{\partial V^k(x)}{\partial x}\right)$ for $\forall x \in \Omega$, resulting in greedy policies.

$$u^{k+1} = -\frac{1}{2} R^{-1} g^T(x) \frac{\partial V^k(x)}{\partial x} \quad (21)$$

$$w^{k+1} = \frac{1}{2\gamma^2} k^T(x) \frac{\partial V^k(x)}{\partial x} \quad (22)$$

Iterate the above three steps until the value function meets some termination conditions, e.g., $|V^{k+1}(x) - V^k(x)| \leq \epsilon$, where $\epsilon$ is a positive definite tolerance.

In order to ensure that policy evaluation step (20) holds or roughly holds for $\forall x \in \Omega$, it may need internal iterations and increased computation. In addition, the value function and policies are usually approximated by linear method, which requires manually designed feature functions, such as polynomial bases with limited approximation ability. Furthermore, it is more challenging to design features for high-dimensional and large-scale problems. Moreover, the initial control policy $u^0$ is required to be stable, and updating policies in (21) and (22) to generate greedy ones requires that the dynamic is input-affine in many previous studies.

*B. Fundamental of Ternary Policy Iteration Algorithm*

The proposed TPI algorithm effectively makes up for the above deficiencies. Three NNs are employed to approximate value function $V(x; \omega)$, control policy $u(x; \theta)$ and disturbance policy $w(x; \eta)$, which are called value network, policy network and disturbance network separately in this study. The parameters of these NNs are denoted as $\omega$, $\theta$ and $\eta$, respectively, and their initial values are set to zero. Note that the NN approximators employed here can also be replaced by polynomial bases to enhance the versatility of the algorithm. Instituting three approximate functions to the Hamiltonian in (7) gives the following approximate Hamiltonian

$$\begin{aligned} H(x, \theta, \eta, \omega) = l(x, u(x; \theta), w(x; \eta)) \\ + \frac{\partial V(x; \omega)}{\partial x^T} \big(f(x) + g(x) u(x; \theta) + k(x) w(x; \eta)\big) \end{aligned} \quad (23)$$

The existing algorithms need to update value function, control policy and disturbance policy for each state of the state set $\Omega$ at the same time, which is impractical. The learning process of our algorithm also contains the above three phases, but value function and policies are updated by diminishing the designed loss functions using the gradient descent (GD) method. The operation of each approximate function in each phase is described below.

| Algorithm 1 Ternary Policy Iteration |
|---|
| Initialization: Zero-initialize parameters of control policy $u(x;\theta^0)$ and disturbance policy $w(x;\eta^0)$, i.e. $\theta^0=0$, $\eta^0=0$. Set $k=0$ |
| Parameters: Learning rates of the value function, control policy and disturbance policy are $\alpha_\omega$, $\alpha_\theta$ and $\alpha_\eta$ |
| 0. Apply control policy $u(x;\theta^k)$ and disturbance policy $w(x;\eta^k)$ to multiple agents with the same dynamic system, and update data set $\mathcal{D}$ <br> 1. For control policy $u(x;\theta^k)$ and disturbance policy $w(x;\eta^k)$, update value function $V(x;\omega^k)$ $$\omega \leftarrow \omega - \alpha_\omega \nabla_\omega L_\omega$$ 2. Given value function $V(x;\omega^{k+1})$, update control policy $u(x;\theta^k)$ and disturbance policy $w(x;\eta^k)$ $$\theta \leftarrow \theta - \alpha_\theta \nabla_\theta L_\theta$$ $$\eta \leftarrow \eta - \alpha_\eta \nabla_\eta L_\eta$$ 3. Go back to 0 and $k \leftarrow k+1$ |

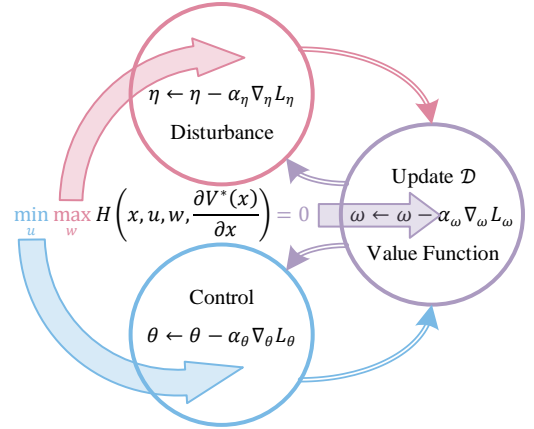

Fig. 1 Ternary Policy Iteration Algorithm

*1) Value Function Update Phase*

In order to avoid operating all states in the entire state set $\Omega$ or performing iterations in value function update phase, define the value loss $L_\omega$ as the expectation of the absolute value of the approximate Hamiltonian in a generated state set $\mathcal{D}$:

$$L_\omega(\omega^k, \theta^k, \eta^k) = \mathbb{E}_{x \in \mathcal{D}}[|H(x, \theta^k, \eta^k, \omega^k)|] \quad (24)$$

Instead of making the Hamiltonian equal to zero in the entire state set $\Omega$, the value function update phase gradually reduces the expectation value of $|H(x, \theta^k, \eta^k, \omega^k)|$ in the generated state set $\mathcal{D}$ by applying GD to the value loss $L_\omega$. The parameters $\omega$ of the value function is updated via

$$\omega^{k+1} = \omega^k - \alpha_\omega \nabla_{\omega^k} L_\omega(\omega^k, \theta^k, \eta^k) \quad (25)$$

$\mathcal{D} \subseteq \Omega$ is generated by multiple agents, which is continuously updated by policies $u(x;\theta^k)$ and $w(x;\eta^k)$ to improve the stability of the update process.

*2) Control Policy Update Phase*

Similar to the previous phase, the control policy loss $L_\theta$ is defined as the expectation of the approximate Hamiltonian:

$$L_\theta(\omega^{k+1}, \theta^k, \eta^k) = \mathbb{E}_{x \in \mathcal{D}}[H(x, \theta^k, \eta^k, \omega^{k+1})] \quad (26)$$

The greedy control policy can only be obtained for input-affine dynamics by minimizing the Hamiltonian in (21). The minimum value of the Hamiltonian is often tricky to solve explicitly. Therefore, the control policy update phase just reduces the expectation value of $H(x, \theta^k, \eta^k, \omega^{k+1})$ by applying GD to the control policy loss $L_\theta$ to update the parameters $\theta$ of control policy. Thus the TPI algorithm can be applied to non-affine systems, which are unsolvable for existing algorithms.

$$\theta^{k+1} = \theta^k - \alpha_\theta \nabla_{\theta^k} L_\theta(\omega^{k+1}, \theta^k, \eta^k) \quad (27)$$

*3) Disturbance Policy Update Phase*

Similar to the control policy loss $L_\theta$, the disturbance policy loss $L_\eta$ is defined as

$$L_\eta(\omega^{k+1}, \theta^k, \eta^k) = \mathbb{E}_{x \in \mathcal{D}}[-H(x, \theta^k, \eta^k, \omega^{k+1})] \quad (28)$$

The positive definiteness of the second-order partial derivative of $w$ is precisely opposite to that of $u$, which accounts for the opposite update direction of the two policies. By adding the minus sign before $H(x, \theta^k, \eta^k, \omega^{k+1})$, GD method can also be utilized to update the parameters $\eta$ of disturbance policy.

$$\eta^{k+1} = \eta^k - \alpha_\eta \nabla_{\eta^k} L_\eta(\omega^{k+1}, \theta^k, \eta^k) \quad (29)$$

The control policy loss $L_\theta$ and the disturbance policy loss $L_\eta$ are opposite numbers, so the updating process of two policies is simultaneous. However, asynchronous updates are also allowed, which is related to their learning speed.

The pseudo-code of the TPI algorithm is presented in **Algorithm 1**, and corresponding iteration procedure is shown in Fig. 1.

IV. SIMULATION RESULTS

In this section, the TPI algorithm is first applied to a linear plant. In order to compare the solution of the algorithm and that of GARE, polynomials are chosen as feature functions. Then, it is applied to a nonlinear system to demonstrate its robustness, where fully connected networks are used.

*A. Simulation of Linear Plant*

The first simulation utilizes an aircraft system described in [24], whose state $x$ consists of the angle of attack $\alpha$, the rate $q$, and the elevator deflection angle $\delta_e$. The controller of the plant is the elevator actuator voltage $u$ and the disturbance is wind gusts $w$ on the angle of attack $\alpha$. In the utility function $l(x, u, w)$, $Q = I$, $R = I$ and $\gamma = 5$. Set value network as $V(x;\omega) = x^T P x = \omega^T \sigma(x)$, policy network as $u(x;\theta) = \theta^T x$ and disturbance network as $w(x;\eta) = \eta^T x$, where the feature function and weights of value network are

$$\omega = [P_{11} \quad 2P_{12} \quad 2P_{13} \quad P_{22} \quad 2P_{23} \quad P_{33}]^T$$
$$\sigma(x) = [\alpha^2 \quad \alpha q \quad \alpha \delta_e \quad q^2 \quad q\delta_e \quad \delta_e^2]^T \quad (30)$$

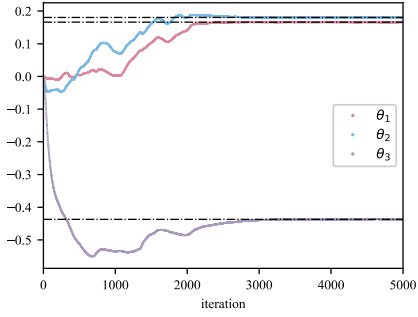

Fig. 5  Weighting values of policy network

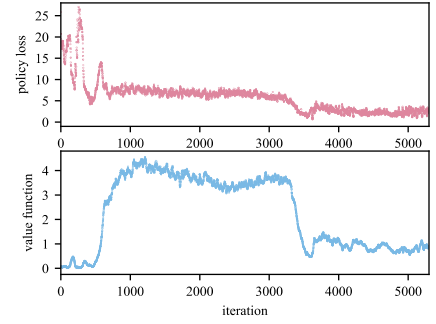

Fig. 2  Learning process of the linear system

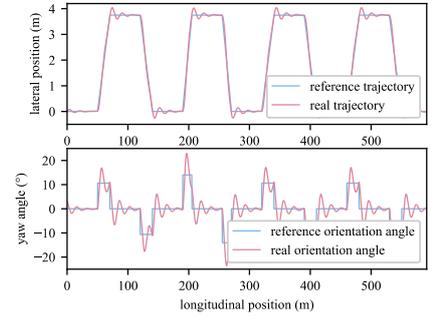

Fig. 3  Tracking effect

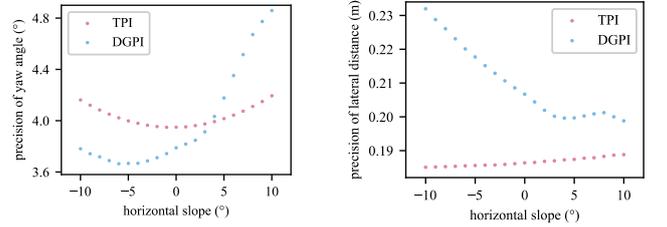

(a) Precision of yaw angle  (b) Precision of lateral distance

Fig. 4  Robustness of TPI

The corresponding GARE (19) can be solved directly, and the robust optimal controller is

$$u^*(x;\theta^*) = \theta^{*T}x$$

$$\theta^* = -PBR^{-1} = [0.1661 \quad 0.1804 \quad -0.4371]^T$$

Set the number of agents as $2^6$. The initial learning rates of the value network, policy network and disturbance network are 0.5, 1.0 and 1.0, and zero-initialization is applied. It can be inferred from Fig. 2 that the weights of policy network converge after 2500 iterations, where three gray lines represent the three components of $\theta^*$. After 5000 iterations, the weights of control policy are obtained as follows

$$\theta = [0.1662 \quad 0.1809 \quad -0.4367]^T$$

The relative error between $\theta$ and the parameters $\theta^*$ of the optimal controller is only around 0.13% in the sense of Euclidean norm. Consequently, the effectiveness of the TPI algorithm in the linear system has been verified.

*B. Simulation of Nonlinear Plant*

Consider a dynamic vehicle model (31) with horizontal slope disturbance, which is an input non-affine nonlinear system [14].

$$\begin{bmatrix} \dot{v}_x \\ \dot{v}_y \\ \dot{\omega}_r \\ \dot{\varphi} \\ \dot{y} \end{bmatrix} = \begin{bmatrix} a_x - \dfrac{F_{yf}\sin\delta}{m} + v_y\omega_r \\ \dfrac{F_{yf}\cos\delta + F_{yr}}{m} - v_x\omega_r \\ \dfrac{aF_{yf}\cos\delta - bF_{yr}}{I_{zz}} \\ \omega_r \\ v_x\sin\varphi + v_y\cos\varphi \end{bmatrix} + \begin{bmatrix} g\sin\varphi \\ g\cos\varphi \\ 0 \\ 0 \\ 0 \end{bmatrix}\sin\beta \quad (31)$$

where $v_x$ is longitudinal velocity, $v_y$ is lateral velocity, $\omega_r$ is yaw rate, $\varphi$ is yaw angle between vehicle and reference trajectory, $y$ is lateral distance between the center of gravity and reference trajectory. The reference trajectory is a periodic double lane change, as shown in Fig. 4. The control input is $u = [\delta \quad a_x]^T$, where $\delta$ is steering angle and $a_x$ is longitudinal acceleration. The disturbance $w$ is a sine function of the horizontal slope $\beta$. The cornering stiffness of the front and rear wheels are $-67104$ N/rad and $-48431$ N/rad, respectively, and the other parameters are taken from [14]. The small-angle approximation is applied to steering angle $\delta$, i.e. $\sin\delta \approx \delta$, $\cos\delta \approx 1$ and $\delta^2/m \approx 0$, to obtain an input-affine nonlinear model. The utility function is defined as

$$l(x,u,w) = 0.5(v_x - 10)^2 + 0.1\omega_r^2 + 36\varphi^2$$
$$+20y^2 + u^T\begin{bmatrix} 0.1 & \\ & 0.3 \end{bmatrix}u - 5^2w^Tw$$

where the desired longitudinal velocity is 10m/s.

To implement the TPI algorithm, three 3-layer fully connected networks are employed, all of which consist of $2^8$ neurons per layer. All the activation functions of three networks are SELU, except that the output layers of the value network, policy network and disturbance network select softplus, tanh and tanh, respectively. The output layer of the policy network multiplies $\begin{bmatrix} \pi/9 & 5 \end{bmatrix}^T$ to adjust the amplitude of the control input. Set the number of agents as $2^{10}$. The initial learning rate of the three networks is $10^{-4}$. Three networks are zero-initialized, and Adam method is implemented to update the networks.

It can be seen from Fig. 3 that both value network and policy network gradually converge in 5000 iterations. Choose one of the policy networks as control policy, and set horizontal slope disturbance as $w = \sin 1°$. The trajectory tracking effect of the vehicle is shown in Fig. 4.

In order to further test and verify the robustness of the TPI algorithm, a comparison simulation with the DGPI algorithm in [14] is performed. The dynamic parameters, utility function, network parameters, algorithm framework and learning process are the same as the TPI, apart from the omission of disturbance update phase. The horizontal slope angle ranges from $-10°$ to $10°$. To compare the control effect of two algorithms, we first define the control precision of yaw angle and lateral position as

$$I_\varphi = \sqrt{\frac{\sum_{i=1}^{N}\left(\varphi(i) - \varphi_{ref}(i)\right)^2}{N}}$$
$$I_y = \sqrt{\frac{\sum_{i=1}^{N}\left(y(i) - y_{ref}(i)\right)^2}{N}} \quad (32)$$

where $\varphi_{ref}$ and $y_{ref}$ are the yaw angle and lateral distance of the reference trajectory.

Fig. 5 shows the comparison results of the control precision of the two algorithms. The extents of the control precision $I_\varphi$ of yaw angle of the TPI is from 3.95° to 4.19°, while that of the DGPI is from 3.66° to 4.86°. So the ranges of $I_\varphi$ of the two algorithms are 0.24° and 1.20° separately. Similarly, the ranges of $I_y$ of the two algorithms are 0.004m and 0.233m, respectively. It can be found that the control precision of the TPI algorithm changes only slightly within the range of horizontal slope compared with the DGPI algorithm. Therefore, the TPI algorithm has a better robust performance in the nonlinear case.

## V. Conclusion

The proposed ternary policy iteration (TPI) algorithm is able to solve robust control problems with nonlinear dynamics. Three loss functions which are designed as the expectation of the Hamiltonian in the generated state set can prevent operating all the states in the entire state set. The corresponding HJI equation of zero-sum game is solved iteratively via decreasing the designed loss functions in three updating phases, respectively. Two simulation results demonstrate the effectiveness and robustness of the proposed algorithm. The TPI algorithm can converge to the optimal solution for the linear plant, and has better resistance to disturbances compared with the DGPI for the nonlinear plant.

## References

[1] G. Zames, "Feedback and optimal sensitivity: Model reference transformations, multiplicative seminorms, and approximate inverses," *IEEE Trans. Automat. Contr.*, vol. 26, no. 2, pp. 301-320, 1981.

[2] B. Francis and G. Zames, "On $H_\infty$-optimal sensitivity theory for SISO feedback systems," *IEEE Trans. Automat. Contr.*, vol. 29, no. 1, pp. 9-16, 1984.

[3] B. A. Francis and J. C. Doyle, "Linear control theory with an $H_\infty$ optimality criterion," *SIAM J. Control Optim.*, vol. 25, no. 4, pp. 815-844, 1987.

[4] J. Doyle, K. Glover, P. Khargonekar, and B. Francis, "State-space solutions to standard $H_2$ and $H_\infty$ control problems," in *1988 American Control Conf.*, 1988, pp. 1691-1696.

[5] F. Gao, S. E. Li, Y. Zheng, and D. Kum, "Robust control of heterogeneous vehicular platoon with uncertain dynamics and communication delay," *IET Intell. Transport Syst.*, vol. 10, no. 7, pp. 503-513, 2016.

[6] S. E. Li, X. Qin, K. Li, J. Wang, and B. Xie, "Robustness analysis and controller synthesis of homogeneous vehicular platoons with bounded parameter uncertainty," *IEEE/ASME Trans. on Mechatronics*, vol. 22, no. 2, pp. 1014-1025, 2017.

[7] S. E. Li, F. Gao, K. Li, L. Wang, K. You, and D. Cao, "Robust longitudinal control of multi-vehicle systems—A distributed $H_\infty$ method," *IEEE Trans. Intell. Transport. Syst.*, vol. 19, no. 9, pp. 2779-2788, 2017.

[8] J. A. Ball and J. W. Helton, "Nonlinear $H_\infty$ control theory for stable plants," *Math. Control Signals Syst.*, vol. 5, no. 3, pp. 233-261, 1992.

[9] A. J. Van Der Schaft, "L$_2$-gain analysis of nonlinear systems and nonlinear state feedback $H_\infty$ control," *IEEE Trans. Automat. Contr.*, vol. 37, no. 6, pp. 770-784, 1992.

[10] D. Liu, Q. Wei, D. Wang, X. Yang, and H. Li, *Adaptive dynamic programming with applications in optimal control*. Berlin: Springer, 2017.

[11] H. Zhang, D. Liu, Y. Luo, and D. Wang, *Adaptive dynamic programming for control: algorithms and stability*. Springer Science & Business Media, 2012.

[12] M. Abu-Khalaf and F. L. Lewis, "Nearly optimal control laws for nonlinear systems with saturating actuators using a neural network HJB approach," *Automatica*, vol. 41, no. 5, pp. 779-791, 2005.

[13] A. Al-Tamimi, F. L. Lewis and M. Abu-Khalaf, "Discrete-time nonlinear HJB solution using approximate dynamic programming: Convergence proof," *IEEE Trans. Syst., Man, Cybern. B*, vol. 38, no. 4, pp. 943-949, Aug. 2008.

[14] J. Duan, S. E. Li, Z. Liu, M. Bujarbaruah, and B. Cheng, "Generalized policy iteration for optimal control in continuous time," *arXiv preprint arXiv:1909.05402*, 2019.

[15] J. Duan, Z. Liu, S. E. Li, Q. Sun, Z. Jia, and B. Cheng, "Deep adaptive dynamic programming for nonaffine nonlinear optimal control problem with state constraints," *arXiv preprint arXiv:1911.11397*, 2019.

[16] R. S. Sutton and A. G. Barto, *Reinforcement learning: An introduction*. MIT press, 2018.

[17] S. E. Li, *Reinforcement learning and control - Lecture notes*. Tsinghua University, 2019.

[18] Z. Jiang and Y. Jiang, "Robust adaptive dynamic programming for linear and nonlinear systems: An overview," *Eur. J. Control*, vol. 19, no. 5, pp. 417-425, Sep. 2013.

[19] M. Abu-Khalaf, F. L. Lewis and J. Huang, "Policy iterations on the Hamilton-Jacobi-Isaacs equation for $H_\infty$ state feedback control with input saturation," *IEEE Trans. Automat. Contr.*, vol. 51, no. 12, pp. 1989-1995, Dec. 2006.

[20] M. Abu-Khalaf, F. L. Lewis and J. Huang, "Neurodynamic programming and zero-sum games for constrained control systems," *IEEE Trans. Neural Networks*, vol. 19, no. 7, pp. 1243-1252, 2008.

[21] A. Al-Tamimi, F. L. Lewis and M. Abu-Khalaf, "Model-free Q-learning designs for linear discrete-time zero-sum games with application to $H_\infty$ control," *Automatica*, vol. 43, no. 3, pp. 473-481, 2007.

[22] H. Zhang, Q. Wei and D. Liu, "An iterative adaptive dynamic programming method for solving a class of nonlinear zero-sum differential games," *Automatica*, vol. 47, no. 1, pp. 207-214, 2011.

[23] H. Wu and B. Luo, "Neural network based online simultaneous policy update algorithm for solving the HJI equation in nonlinear $H_\infty$ control," *IEEE Trans. Neural Netw. Learn. Syst.*, vol. 23, no. 12, pp. 1884-1895, 2012.

[24] H. Modares, F. L. Lewis and Z. Jiang, "$H_\infty$ tracking control of completely unknown continuous-time systems via off-policy reinforcement learning," *IEEE Trans. Neural Netw. Learn. Syst.*, vol. 26, no. 10, pp. 2550-2562, 2015.

[25] D. Liu, X. Yang, D. Wang, and Q. Wei, "Reinforcement-learning-based robust controller design for continuous-time uncertain nonlinear systems subject to input constraints," *IEEE Trans. Cybern.*, vol. 45, no. 7, pp. 1372-1385, 2015.

[26] T. Basar and P. Bernhard, $H_\infty$ *optimal control and related minimax design problems: a dynamic game approach*. Springer Science & Business Media, 2008.

[27] F. L. Lewis, D. Vrabie and V. L. Syrmos, *Optimal control*. John Wiley & Sons, 2012.